\documentclass[10pt]{article}
\usepackage[letterpaper]{geometry}
\usepackage{hicss}
\usepackage{times}
\usepackage[none]{hyphenat}
\usepackage{url}
\usepackage{latexsym}
\usepackage{indentfirst}
\usepackage{graphicx}
\graphicspath{{images/}}
\usepackage{graphicx}
\usepackage{algorithm}
\usepackage{algorithmic}
\usepackage{xcolor}
\usepackage{amsmath,amsfonts,amsthm,amssymb,amsopn,bm,mathtools,comment}
\newtheorem{proposition}{Proposition}

\usepackage{amsmath,amsfonts,bm}

\def\1{\bm{1}}

\def\vb{{\bm{b}}}

\def\vp{{\bm{p}}}
\def\vq{{\bm{q}}}

\def\vx{{\bm{x}}}
\def\vy{{\bm{y}}}
\def\vz{{\bm{z}}}

\def\mU{{\bm{U}}}
\def\mV{{\bm{V}}}
\def\mW{{\bm{W}}}

\DeclareMathAlphabet{\mathsfit}{\encodingdefault}{\sfdefault}{m}{sl}
\SetMathAlphabet{\mathsfit}{bold}{\encodingdefault}{\sfdefault}{bx}{n}

\usepackage{subfig}
\usepackage{booktabs}

\title{A Unified Approach to Enforce Non-Negativity Constraint in Neural Network Approximation for Optimal Voltage Regulation}

\author{Jiaqi Wu \\
  Arizona State University \\
  {\underline{jiaqiwu1@asu.edu}} \\  \And
  Jingyi Yuan \\
  Arizona State University \\
  {\underline{jyuan46@asu.edu} } \\ \And
  Yang Weng \\
  Arizona State University \\
  {\underline{yweng2@asu.edu} } \\ \And
  Guangwen Wang\\
  Arizona State University \\
  {\underline{gwang114@asu.edu} } \\
 }
\date{}

\begin{document}
\maketitle
\begin{abstract}
Power system voltage regulation is crucial to maintain power quality while integrating intermittent renewable resources in distribution grids. However, the system model on the grid edge is often unknown, making it difficult to model physical equations for optimal control. Therefore, previous work proposes structured data-driven methods like input convex neural networks (ICNN) for ``optimal" control without relying on a physical model. While ICNNs offer theoretical guarantees based on restrictive assumptions of non-negative neural network (NN) parameters, can one improve the approximation power with an extra step on negative duplication of inputs? We show that such added mirroring step fails to improve accuracy, as a linear combination of the original input and duplicated input is equivalent to a linear operation of ICNN's input without duplication. While this design can not improve performance, we propose a unified approach to embed the non-negativity constraint as a regularized optimization of NN, contrary to the existing methods, which added a loosely integrated second step for post-processing on parameter negation. Our integration directly ties back-propagation to simultaneously minimizing the approximation error while enforcing the convexity constraints. Numerical experiments validate the issues of the mirroring method and show that our integrated objective can avoid problems such as unstable training and non-convergence existing in other methods for optimal control. 
\end{abstract}

\subsubsection*{Keywords:}

Input convex neural network, power system voltage regulation, input duplication, constraint integration, back-propagation

\section{Introduction}
Integrating distributed energy resources (DERs) requires efficient voltage regulation methods on the grid edge, maintaining power quality \cite{wu2022spatial}. The reason is that the growing integration of intermittent photovoltaics, electric vehicles, and volatile loads elevate voltage levels and create voltage oscillations \cite{wu2023learn,murray2021voltage}. Therefore, effective voltage regulation methods are essential for the stable and reliable operation of power grids \cite{mirafzal2020grid}. Edge grids typically utilize controllers, such as inverters equipped with decentralized resources, to adjust reactive (or active) power injections. The objective is to maintain operational voltage levels within standard range \cite{srivastava2023voltage}.

Some of the traditional approaches for voltage regulation are based on the physical model of optimal power flow (OPF), which aims to find an optimal adjustment of reactive power. With the governing power flow (PF) equations describing the couplings among physical variables between measurements and controllers, such model-based approaches are reliable even if the operating point changes occasionally. However, they face significant challenges of non-convex and NP-hard voltage control problems due to the nonlinearity of PF modeling \cite{farivar2013branch}. Therefore, past works propose various convexification strategies, such as linearization \cite{zhu2015fast}, semidefinite and second-order cone programming \cite{zhang2014optimal,farivar2012optimal}, and Lagrangian dual problem \cite{lavaei2011zero}.

\begin{figure*}[h!]
    \centering
    \includegraphics[width=\textwidth]{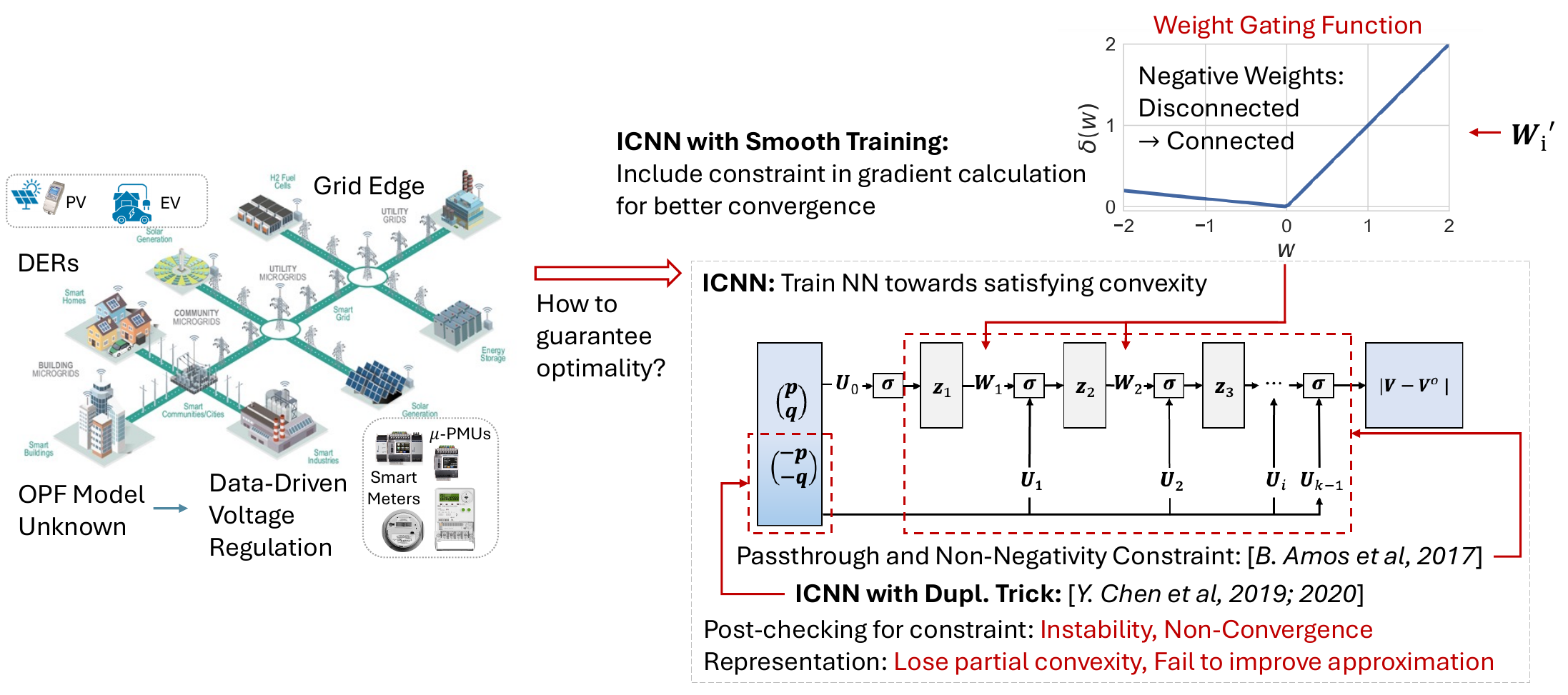}
    \caption{Overview of proposed smooth training in the training of convexified data-driven voltage regulation.}
    \label{fig:bigpic}
\end{figure*}

However, traditional model-based methods assume complete knowledge of PF physics, which is often unavailable in edge grids due to unreported topology changes and delayed information on system upgrades \cite{yuan2022intrinsic}. Such a problem is shown on the left side of Fig. \ref{fig:bigpic}. The unknown system parameters make it infeasible to apply physical model-based control strategies on distribution grid edges \cite{nouri2021voltage, wang2023reinforcement}. Recognizing that the models are implicitly embedded in data, recent work has shifted toward data-driven methods for control due to new communication infrastructure \cite{yuan2022intrinsic}. Subsequently, data-driven PF models are developed for voltage regulation \cite{li2023distribution}.

Neural networks (NN), in particular, have been proposed as nonlinear surrogates to approximate PF equations \cite{choi2021convex,wang2019integrating}.  While they enable the formulation of voltage control optimization without a physical model, NNs' black-box nature aggravates the difficulty of finding an optimal solution. To achieve optimum voltage regulation, some approaches use traditional convex relaxation techniques for OPF formulation to combine with data-driven methods \cite{xu2019data, ayyagari2019artificial}. As the relaxation error and approximation errors progressively increase, one promising approach involves embedding convexity directly into the data-driven model using input convex neural networks (ICNNs)  \cite{chen2020data}.
ICNNs ensure a convex relationship from input to output via constraints on NN architectures \cite{amos2017input}. Thus, can we make ICNN facilitate consistent identification of minimum voltage deviations by optimizing reactive power injections?


Previous works on power system control problems apply ICNNs for complex operations, such as nonlinear control behaviors described by partial differential equations \cite{Wu2023Transient}. The global optimality resulting from their convexity constraints is assumed to provide stable solutions in model predictive control \cite{lawrynczuk2022input}. 
However,  \cite{Makkuva2020Optimal} points out that the accuracy of ICNNs 
requires further validation. A similar concern is echoed in works like \cite{Su2023Analytic, Wu2023Transient} where ICNNs are shown to evaluate the transient stability of power systems but may exhibit higher errors on complex dynamics. Moreover, the convexity constraints on non-negative weights simultaneously limit NNs' representation power. ICNNs require more iterations during training to achieve convergence \cite{Makkuva2020Optimal, Yang2021Optimization-based}.

As Fig. \ref{fig:bigpic} shows in the lower right,  the key design of the original ICNN paper is on embedding ``passthrough'' connections to provide an additional linear path for representation complement \cite{amos2017input}. Past work attempts to improve the representation power by introducing the mirroring method to expand input variables and corresponding parameter space \cite{chen2018optimal}.
However, we show that such a design retains an equivalent representation with redundant parameters and causes a loss of convexity in the original mapping. 
Due to such a fact, the mirroring method based on duplication is infeasible for resolving the trade-off of convexification and approximation efficiency in NNs. 

Beyond duplication, we found out that existing methods and their codes implement the method in two stages. One stage is to train the NN according to the typical setup. The second stage modifies the negative coefficient to be non-negative. Solving the ICNN problem in a two-stage setup will cause training instability and non-convergence problems. 
In particular, previous ways of training ICNNs separately update the model with respect to minimizing loss and satisfying convexity constraints. Such a post-processing mechanism frequently oscillates weights during iterations and hinders convergence. Instead, we unify the two steps into a single NN optimization by embedding the post-processing step into the back-propagation process. This design leads to a gated function inside the ICNN training process to enforce non-negative weights, as shown in Fig. \ref{fig:bigpic}. The key outcome of the gate design is mitigating the vanishing gradient problem during loss minimization. 
Numerical results on different ICNN implementations show the restriction of convexity on representation power and demonstrate the effectiveness of our smooth convexification technique using the proposed gate function. 

The remainder of the paper is structured as follows. Section \ref{sec:voltage_regulation} presents the mathematical modeling. Section \ref{sec:method} shows why the mirroring method does not work and how to design a gate function to integrate the non-negativity constraints into back-propagation. Section  \ref{sec:experiment} validates our claims, and Section \ref{sec:conclusion} concludes the paper.

\section{Data-driven Voltage Regulation with Convexity} \label{sec:voltage_regulation}

Classic voltage regulation formulates a model-based OPF optimization problem, using power flow equations as constraints to describe the coupling among physical variables in the grid. However, data-driven voltage regulation methods have been developed due to the unobservability of the line connections and parameters in the grid. One such data-driven method aims to learn a convex mapping from the objective function to the variables to preserve the optimality preferred by power engineers. Such a series of work leverages the convex property of the ICNN.

\subsection{Classic Model-Based Voltage Regulation}

In classical settings, voltage regulation is formulated as an OPF problem. The optimization's objective function aims to adjust voltages to their desired operational levels \cite{zhang2013local}. Furthermore, OPF formulations incorporate constraints to describe system operational limits, categorizing these approaches as model-based. For example, \cite{farivar2011inverter} elaborates these constraints within the branch flow model framework, known as the DistFlow \cite{baran1989optimal}, in distribution circuits for Volt/VAR control. The DistFlow equations are
\begin{subequations} \label{eq:distflow}
\begin{align}
    P_{ij} &= \sum_{k:(j,k) \in E} P_{jk} + r_{ij}l_{ij} + p_j , \\
    Q_{ij} &= \sum_{k:(j,k) \in E} Q_{jk} + x_{ij}l_{ij} + q_j , \\
    v_j &= v_i - 2(r_{ij}P_{ij} + x_{ij}Q_{ij}) + (r^2_{ij} + x^2_{ij})l_{ij} , \\
    \forall & (i,j) \in E : l_{ij} = \frac{P^2_{ij} + Q^2_{ij}}{v_i} , \label{eq:distflow_4}
\end{align}
\end{subequations}
where $I_{ij}$, $P_{ij}$, and $Q_{ij}$ are current, active power, and reactive power from bus $i$ to bus $j$, respectively. Moreover, bus voltage magnitude $v_i = V_i^2 $, and line current flow $l_{ij} = I_{ij}^2$. $r_{ij}$ and $x_{ij}$ are resistance and reactance of line $(i, j)$. The DistFlow model provides a comprehensive physical representation of distribution grids across a range of operational conditions, thereby facilitating accurate analysis and optimization of voltages with generalization.

To ensure the solution remains feasible and efficient under operational constraints, traditional methods implement convexity relaxation techniques for the OPF problem. The relaxation of DistFlow model involves replacing \eqref{eq:distflow_4} as follows:
\begin{align}
    \forall (i,j) \in E : l_{ij} \geq \frac{P^2_{ij} + Q^2_{ij}}{v_i}, \label{eq:distflow_5}
\end{align}
thereby formulating the voltage regulation OPF as a second-order cone program (SOCP). 
Following the requirement in \cite{low2014convex}, this relaxation is exact, and the new optimum remains optimal for the original OPF without relaxation \cite{farivar2013branch}.

Although convex relaxation enables traditional model-based methods to solve voltage regulation OPF problems effectively, a significant challenge is limited system knowledge. Specifically, distribution systems often lack comprehensive data on system topology and line parameters. The topology information, detailing the connections between buses $i$ and $j$, and line parameters, which include impedance $r_{ij}$ and $x_{ij}$ of the distribution lines, are essential for precise modeling. Without knowing the topology and line parameters, the distribution grid cannot be formulated explicitly.

To tackle the unobservability, \cite{chen2018optimal} uses a machine learning model to substitute the constraints based on the DistFlow framework to model the power system. In this method, the objective function aims to directly minimize the discrepancy between the measured voltage magnitude $V_i$ and its reference value $V_i^o$, and the NN approximates the mapping from power injection to the voltage discrepancy, which is in the same direction as the DistFlow model from power to voltage. While NN models exhibit universal approximation capabilities, their inherent black-box nature complicates performance assurance in specific optimization tasks. An ICNN is a required approximator to ensure convexity under unobservability. The data-driven voltage regulation OPF can be formally defined as
\begin{align}
    \min_{\vq} \quad & \sum_{i=1}^{s} a_i |V_i - V_{i}^o|  \\
    \text{subject to: } & \vq_{\min} \leq \vq \leq \vq_{\max} ,  \\
    & |\mV - \mV^o| = f(\vp, \vq),\label{eq:opf_icnn}
\end{align}
where $f(\cdot)$ is the ICNN, $\vp = [p_1, p_2, \cdots, p_i]^\top$, $\vq = [q_1, q_2, \cdots, q_i]^\top$, $\mV = [V_1, V_2, \cdots, V_i]^\top$, $\mV^o = [V_1^o, V_2^o, \cdots, V_i^o]^\top$, and $(\vp, \vq)$ denotes the vector concatenation for real and reactive power injection variables. Moreover, system operators can adjust the scaling factor $a_i$ in the objective function. Specifically, the ICNN model learns a convex function for the objective function in terms of the optimization variables, thereby benefiting the finding of the optimum for the optimization.

\subsection{Theoretical Guarantee of Input Convex Neural Networks for Voltage Regulation}

\begin{figure}[h!]
    \centering
    \subfloat[The structure of the ICNN. \label{fig:icnn}]{\includegraphics[width=\columnwidth]{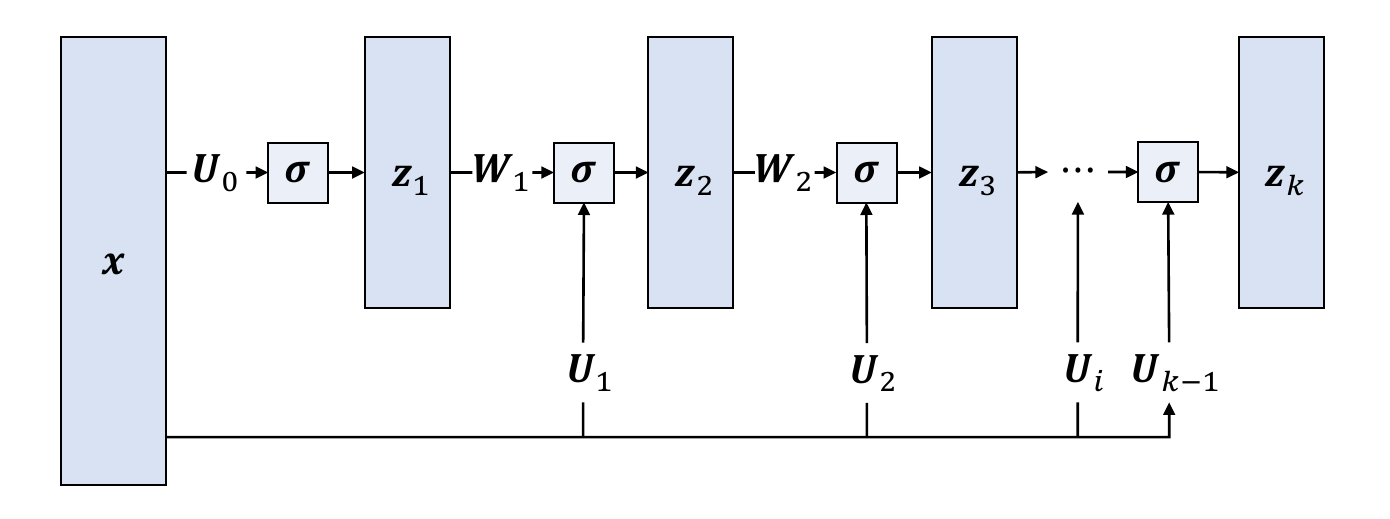}}\\
    \subfloat[Non-decreasing convex activation functions. \label{fig:activation}]{\includegraphics[width=0.9\columnwidth] {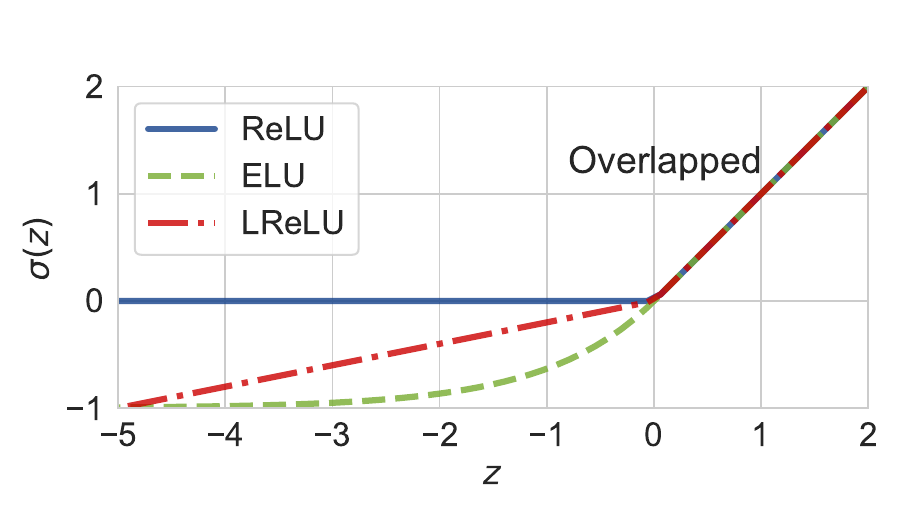}}
    \caption{The ICNN is convex from the input to the output because of the non-negative weights and the convex and non-decreasing activation functions.}
    \label{fig:intro_icnn}
\end{figure}

\paragraph{Convexity Guarantee}
The ICNN is first proposed in \cite{amos2017input}. Given a fully connected $k$-layer NN, an ICNN re-constructs it as a convex function to the inputs, as shown in Figure \ref{fig:icnn}. The mathematical expression of the ICNN is
\begin{align}
    \vz_1 &= \sigma_0 (\mU_0 \vx + \vb_0), \\
    \vz_{i+1} &= \sigma_i (\mW_i \vz_i + \mU_i \vx + \vb_i),
\end{align}
where $\vz_i$ denotes the output of the $i$-th hidden layer in the NN, $\mW_{1:k}$ and $\mU_{0:k}$ are the parameters of the fully connected layers and the ``passthrough'' layers, respectively, and $\sigma_i$ is the activation functions. The convexity from the input $\vx$ to the output $\vy$ is achieved following Proposition \ref{prop:icnn}.

\begin{proposition} \label{prop:icnn}
The neural network is convex from the input to the output, given that all weights in $\mW_{1:k-1}$ are non-negative, and all activation functions $\sigma(\cdot)$ are convex and non-decreasing.
\end{proposition}

The proof of Proposition \ref{prop:icnn} follows the operations that preserve convexity mentioned in \cite{boyd2004convex}. Initially, a non-negative weighted sum of convex functions remains convex. Furthermore, for a function composition $\zeta(\vx) = \mu(\rho(\vx))$, $\zeta$ is convex if $\mu$ is convex and non-decreasing, and $\rho$ is convex.  


The requirement of the non-decreasing convex activation functions is not restrictive. We can choose from some popular activation functions many options, including the rectified linear unit (ReLU) \cite{nair2010rectified}, the leaky rectified linear unit (LReLU) \cite{maas2013rectifier}, and the exponential linear unit (ELU) \cite{clevert2015fast} in Fig. \ref{fig:activation}. These options are commonly used in the computer science community and have been proven effective in various machine-learning applications.

\paragraph{Approximation Guarantee}

\begin{figure}
    \centering
    \includegraphics[width=0.8\columnwidth]{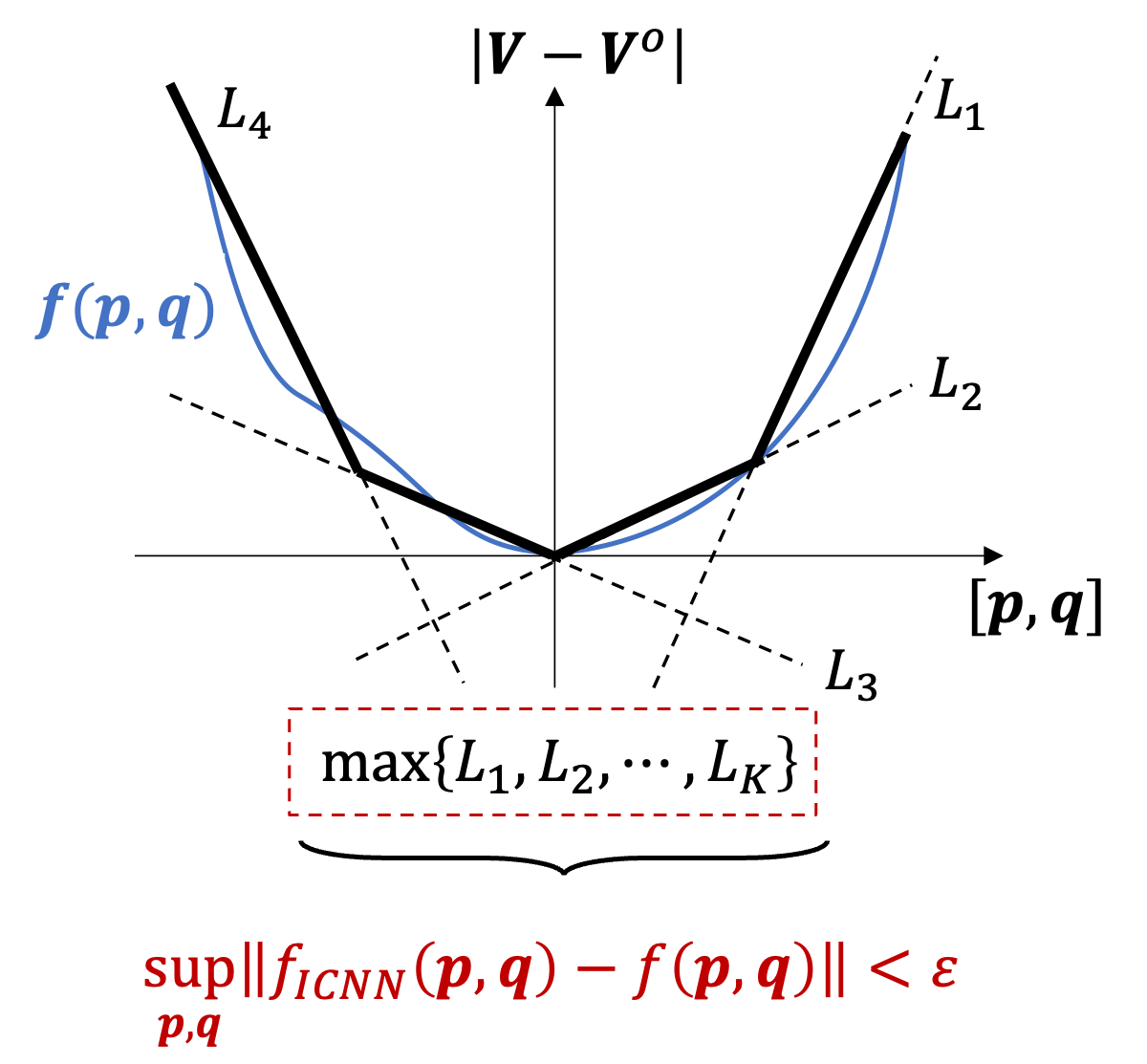}
        \caption{Convex functions can be represented as maximizing a group of affine functions for piecewise linear approximation.}
    \label{fig:approx}
\end{figure}

The ICNN is able to approximate any convex function because of the fact that convex piecewise linear functions can be represented as a maximum of affine functions \cite{wang2004general}. The ICNN, $|\mV - \mV^o| = f(\vp, \vq)$, can be approximated by convex piecewise linear functions, represented as $\max \{ L_1, L_2, \cdots, L_k \}$, as shown in Fig. \ref{fig:approx}.

There are two requirements in the ICNN, which are special activation functions and non-negativity constraints. Commonly used activation functions, such as ReLU and its variants like leaky ReLU, satisfy both convexity and monotonicity, ensuring nonlinearity for universal NN approximation. However, constraining weights to the positive range limits training flexibility and representation efficiency. For example, \cite{chen2018optimal} uses a processing mechanism following standard NN training, replacing negative weights iteratively, which can cause oscillations in gradients and weights. Alternatively, others embed a gating function into the NN architecture to constrain weights, such as the ``clamp function'' in PyTorch. This function clamps all elements in input $x$ into the range $[x_{\min}, x_{\max} ]$. Both strategies are separated from the back-propagation in training, bringing challenges to the convergence of the ICNN. To avoid oscillations in training, we propose implementing the constraint in parallel with training, providing a satisfactory guarantee of a hard constraint for input convexity.



\section{Analyzing Mirroring Strategy and Our Proposed Method} \label{sec:method}

To guarantee the convexity of the ICNN, traditional methods apply a non-negativity constraint. The use of this constraint maintains the convexity of the ICNN, but it also increases the difficulty of achieving training convergence. This section first shows that a trick used in the past to increase representation power is theoretically ineffective and harms training efficiency. Then, we propose an integrated training method to include the constraint in the back-propagation, thereby improving training performance. 

\subsection{Inefficiency of the Duplication Trick in ICNN} \label{sec:inefficiency}

Due to the non-negativity constraint on the weights, the ICNN loses significant representation power despite the linear mapping of the ``passthrough'' layers $\mU_{0:k-1}$ being designed to mitigate this issue. \cite{chen2018optimal} uses a duplicate of $\vx$ to improve the representation power of the basic ICNN. The negative weights of $\mU_{0:k-1}$ in \cite{chen2018optimal} are set to zero, and their negations are set as the weights for corresponding $-\vx$. This way, the forward calculation in the new structure has the same result as the calculation in the original ICNN structure. For instance, consider an input pair $[x_1, x_2]^T$ and the corresponding weights pair $[w_1, w_2]$, where $w_2 < 0$. Following this method, the new weights pair becomes $[w_1, 0, 0, -w_2]^T$. With the combination of the original inputs and mirroring inputs $[x_1, x_2, -x_1, -x_2]$, the result is $(w_1x_1 + w_2x_2)$, which equals the inner product of $[w_1, w_2]$ and $[x_1, x_2]$.


However, in practice, this duplication will not work as expected. The ``passthrough'' layers $\mU_{0:k-1}$ are directly connected to the input. Without constraints on these layers, negative values are allowed and do not compromise the convexity of the ICNN. Consequently, the linear mapping of $\mU_{0:k-1}$ without constraints is equivalent to any other linear transformation in the network, including the negation trick applied to $\mU_{0:k-1}^{(+)}$, the weights of the original input, and $\mU_{0:k-1}^{(-)}$, the weights of the negation of the original input. 

From the perspective of weights updating, the network will reset all weights of $-\vx$ after each iteration, so these weights $\mU_{0:k-1}^{(-)}$ are only considered in the forward calculation, not in the back-propagation. Moreover, since all the weights of $\vx$, i.e., $\mU_{0:k-1}^{(+)}$, are non-negative after each iteration, the negation of their negative value, which is the weights in $\mU_{0:k-1}^{(-)}$ will be small in the new iteration and to the final iteration.

\subsection{Improve the Training Efficiency via Constraint Integration in Back-Propagation} \label{sec:smooth_training}
While Proposition \ref{prop:icnn} lists sufficient conditions for constructing an NN with input convexity, the existing method uses post-processing on non-negative weights after training a regular NN. Such an iterative method can be viewed as adding constraints on standard NN parameters in a second stage. 
As the non-negativity constraint is not integrated with the objective, the back-propagation process will be updated iteratively with an unaligned clamp function, leading to gradient vanishing and training instability.

\begin{figure}[h!]
    \centering
    \includegraphics[width=\columnwidth]{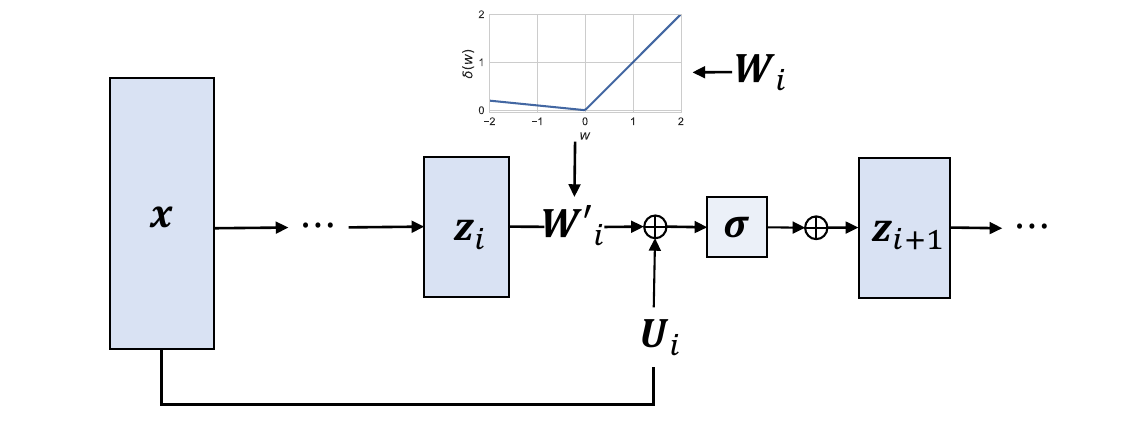}
    \caption{The smooth training via the weight gating design includes the gradient calculation of the non-negativity constraint in the training.}
    \label{fig:smooth_training}
\end{figure}

To boost the training efficiency for better approximation, 
we propose an integrated objective to include the non-negativity constraints in the back-propagation. Specifically, our idea involves a weight gating function as the constraint in each layer of the NN. By doing so, the gradient of the constraint will be calculated during back-propagation, as shown in Fig. \ref{fig:smooth_training}.

We use one example layer to illustrate how to integrate the gradient of the constraint into the training. This layer in the basic ICNN with two parameters can be written as
\begin{align}
z_{i+1} & = \sigma (w_{i, 1} \cdot z_{i, 1} + w_{i,2} \cdot z_{i, 2} + b_i), \\
w_{i, 1} & = \gamma(w_{i, 1}), \\
w_{i, 2} & = \gamma(w_{i, 2}),
\end{align}
where $\sigma(\cdot)$ is the activation function and $\gamma(\cdot)$ is the non-negativity constraint to limit the weights in $\mW_{1:k-1}$. The gradient of $w_{i,1}$ in this layer can be written as 
\begin{align}
\frac{\partial \mathcal{L}}{\partial w_{i,1}} & = \frac{\partial \mathcal{L}}{\partial \hat{z}_{i+1}} \cdot \frac{\partial \hat{z}_{i+1}}{\partial w_{i,1}},
\end{align}
where $\hat{z}_{i+1}$ is the temporary value of the $z_{i+1}$ in the training.

The constraint will not be included in the gradient calculation. 

To include the gradient of the constraint in the training, we propose to use a new layer structure as
\begin{align}
w'_{i,1} &= \delta(w_{i,1}), \\
w'_{i,2} &= \delta(w_{i,2}), \\
z_{i+1} & = \sigma (w'_{i, 1} \cdot z_{i, 1} + w'_{i,2} \cdot z_{i, 2} + b_i),
\end{align}
where $\sigma(\cdot)$ is the activation function and $\delta(\cdot)$ is the weight gating function proposed in this paper. The gradient of $w_{i,1}$ in this layer can be written as 
\begin{align}
\frac{\partial \mathcal{L}}{\partial w_{i,1}} & = \frac{\partial \mathcal{L}}{\partial \hat{z}_{i+1}} \cdot \frac{\partial \hat{z}_{i+1}}{\partial w'_{i,1}} \cdot \frac{\partial w'_{i,1}}{\partial w_{i,1}},
\end{align}
where the last term is the gradient of the non-negativity constraint. However, if we simply use a function of $w' = \max(0, w)$ to constrain the weights, the last term can be zero, and the gradient will vanish. Therefore, we apply a negative slope $s$, which is used for negative input values. The weight gating function can be written as $w' = \max(0, w) + s \cdot \min(0, w)$, where $s \leq 0$ is a hyperparameter that needs to be tuned. Specifically, if a weight $w$ is negative, it will be scaled by $s$ and be converted to a small positive value. If $w$ is positive, it will not be scaled.


\section{Experimental Results} \label{sec:experiment}

This section tests the proposed ideas numerically for two aspects. In  \ref{sec:inefficiency_exp}, we aim to validate our theoretical analysis on the duplication trick in Section \ref{sec:inefficiency}, which may not contribute to ICNN training. 
Based on that, Section \ref{sec:smooth_training_exp} will numerically show how our proposed smooth training algorithm in Section \ref{sec:smooth_training} improves ICNN's training performance.


\subsection{Data Generation}


To prepare data for machine learning models, we use MATPOWER \cite{zimmerman2010matpower} to generate data by running the power flow analysis. MATPOWER is an open-source MATLAB-based power system simulation package that power system researchers widely use. Specifically, we customize an inter-connected $10$-bus case from standard feeders to intuitively demonstrate the inefficiency of the duplication trick in Section \ref{sec:inefficiency_exp}.
For testing cases, we use a $12$-bus feeder \cite{liao2018urban}, a $116$-bus feeder \cite{liao2018urban}, and an Arizona generic utility distribution feeder with 371 buses \cite{wu2022spatial}.
The load profile comes from a $30$-day time-series dataset of every 15 minutes real power consumption $p_i[t]$. For the reactive power, we emulate $q_i[t]$ according to a random lagging power factor $pf_i[t]$, i.e., $pf_i[t] \sim \text{Unif}(0.85, 0.95)$. After running the power flow analysis, we create a refined time-series dataset for this feeder. We use random factors for the unseen operational points to scale the $p_i[t]$ and run the simulation. 

\subsection{The Duplication Trick Will Not Improve the Training Efficiency}
\label{sec:inefficiency_exp}
\begin{figure}[h!]
    \centering
    \includegraphics[width=\columnwidth]{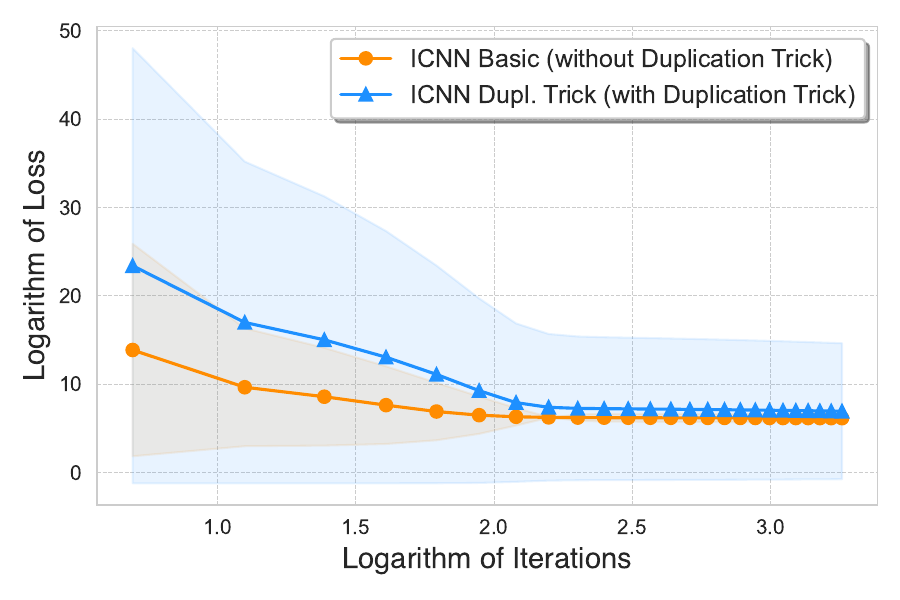}
    \caption{With a similar number of parameters in the models, the basic ICNN and the ICNN with the duplication trick achieve similar training losses.}
    \label{fig:compare_dupilcation_same_para}
\end{figure}

To fairly compare the approximation capability, we assign a basic ICNN with a similar number of trainable parameters, denoted as ICNN Basic, and compare to the ICNN with the duplication trick, denoted as ICNN Dupl. Trick. All experiments are simulated $20$ times under different random seeds on the 10-bus case. We use different metrics to show the difference between the basic ICNN and the ICNN with the duplication trick, including training time, training loss, and mean absolute percentage error (MAPE). The two models have a similar number of trainable parameters, so they theoretically have similar training performance. Since the training loss varies significantly on a wide range, it is hard to observe the entire curve and training details at the same time. Therefore, we convert the loss-iteration plot to a log-log graph to better observe the training performance. From Fig. \ref{fig:compare_dupilcation_same_para}, we observe that using the duplication method has worse performance than the method without mirroring the negation of inputs. Moreover, the MAPE error of the duplication method is slightly larger consistently in Table \ref{tab:duplication}. However, the small difference indicates that the duplication trick does not significantly improve the representation power, which is consistent with the expectation. On the other side, the mirroring of inputs introduces more parameters into the model by extending the size of the input vector. 

\begin{figure}[h!]
    \centering
    \includegraphics[width=\columnwidth]{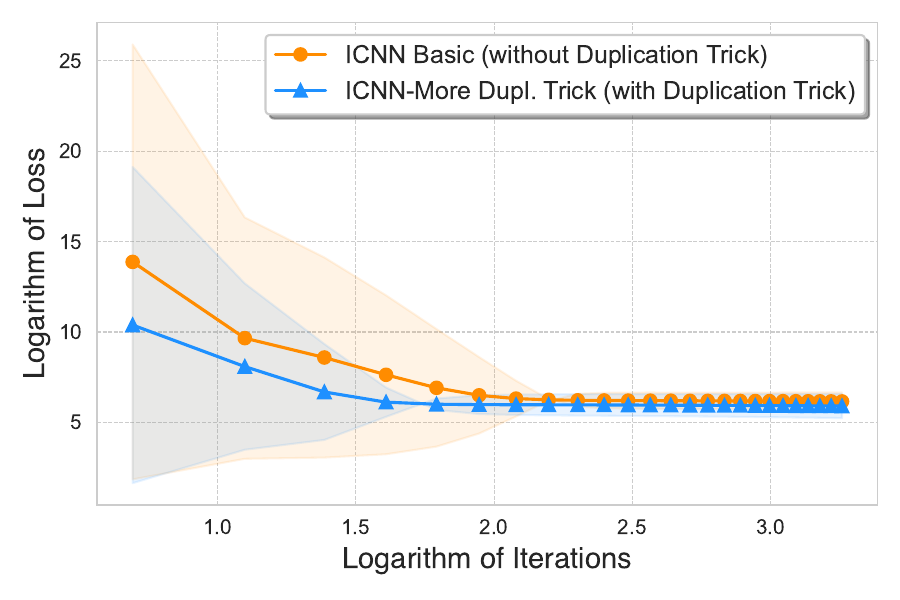}
    \caption{With the same number of hidden neurons in each layer, the ICNN with the duplication trick achieves a smaller training loss but requires more computation time per iteration.}
    \label{fig:compare_dupilcation_same_neuron}
\end{figure}

A natural question is when such a duplication method improves performance. To answer this, we keep the number of hidden neurons in the mirroring method the same as in the original method without mirroring, which leads to an increased number of parameters for NN training. Unlike a basic NN, the ``passthrough'' connections from the input to each hidden layer make the number of parameters in the ICNN significantly dependent on the features of the input vectors. Consequently, the mirrored input increases the total number of trainable parameters in the network. Therefore, we create a new case, denoted as ICNN-More Dupl. Trick, by including additional parameters and observe improved performance in Fig. \ref{fig:compare_dupilcation_same_neuron}. By comparing the results from both setups, we conclude that the duplication trick is not the reason for improving the performance, but the increased parameter is the reason for the boost. 
However, we shall point out that increasing the parameters, e.g., making the NN wider for the method without mirroring, can also improve the performance. However, more parameters result in longer training times for each iteration. 

We summarize all the performance records in 
Table \ref{tab:duplication}. In conclusion, 
the results demonstrate that mirroring the inputs does not improve the ICNN's representation capability. 

%

\begin{table}[h!]
    \centering
    \begin{tabular}{c|c|c|c}
    \toprule
         & ICNN & ICNN & ICNN-More  \\
         & Basic & Dupl. Trick & Dupl. Trick  \\
         \midrule
         Size  & [16, 32, 16] & [10, 20, 10] & [16, 32, 16]  \\
         \midrule
         Time  & 15.56s & 15.81s & 18.73s  \\
         \hline
         Loss & 394.89 & 452.23 & 304.38  \\
         \hline
         MAPE  & 9.1$\%$ &  9.4$\%$ & 7.8$\%$  \\
         \bottomrule
    \end{tabular}
    \caption{The comparison of the ICNN with or without the duplication trick under different experiment setups.}
    \label{tab:duplication}
\end{table}

\subsection{Improved Performance with Back-Propagation Integration} \label{sec:smooth_training_exp}

To compare the proposed method of integrating the back-propagation with non-negativity constraints, we test the loss of voltage regulation results between the proposed method called smoothing training in the legend. The benchmark method is called ICNN with Post-Check. This is because the past method checks if the weights are positive only after back-propagation as a separate stage. After running $20$ experiments for each model on time-series datasets, we convert the results into a log-log graph for better observation of the differences. 

\begin{figure}[h!]
    \centering
    \includegraphics[width=\columnwidth]{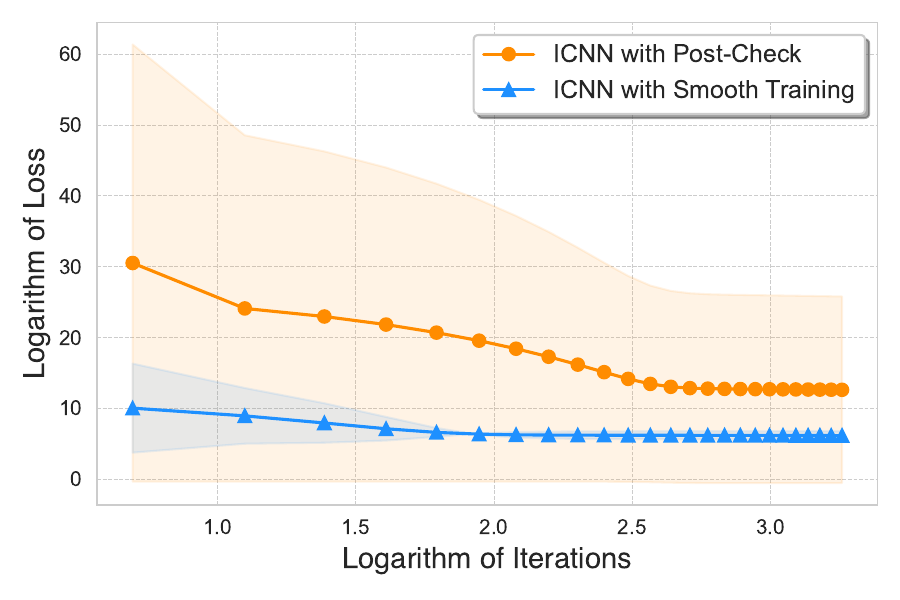}
    \caption{The training error comparison between the basic ICNN and the ICNN with smooth training in the $116$-bus feeder test case. By involving the gradient calculation of the constraint in the training, the ICNN achieves lower training loss with less iterations.}
    \label{fig:compare_smooth}
\end{figure}

\begin{figure}[h!]
    \centering
    \includegraphics[width=\columnwidth]{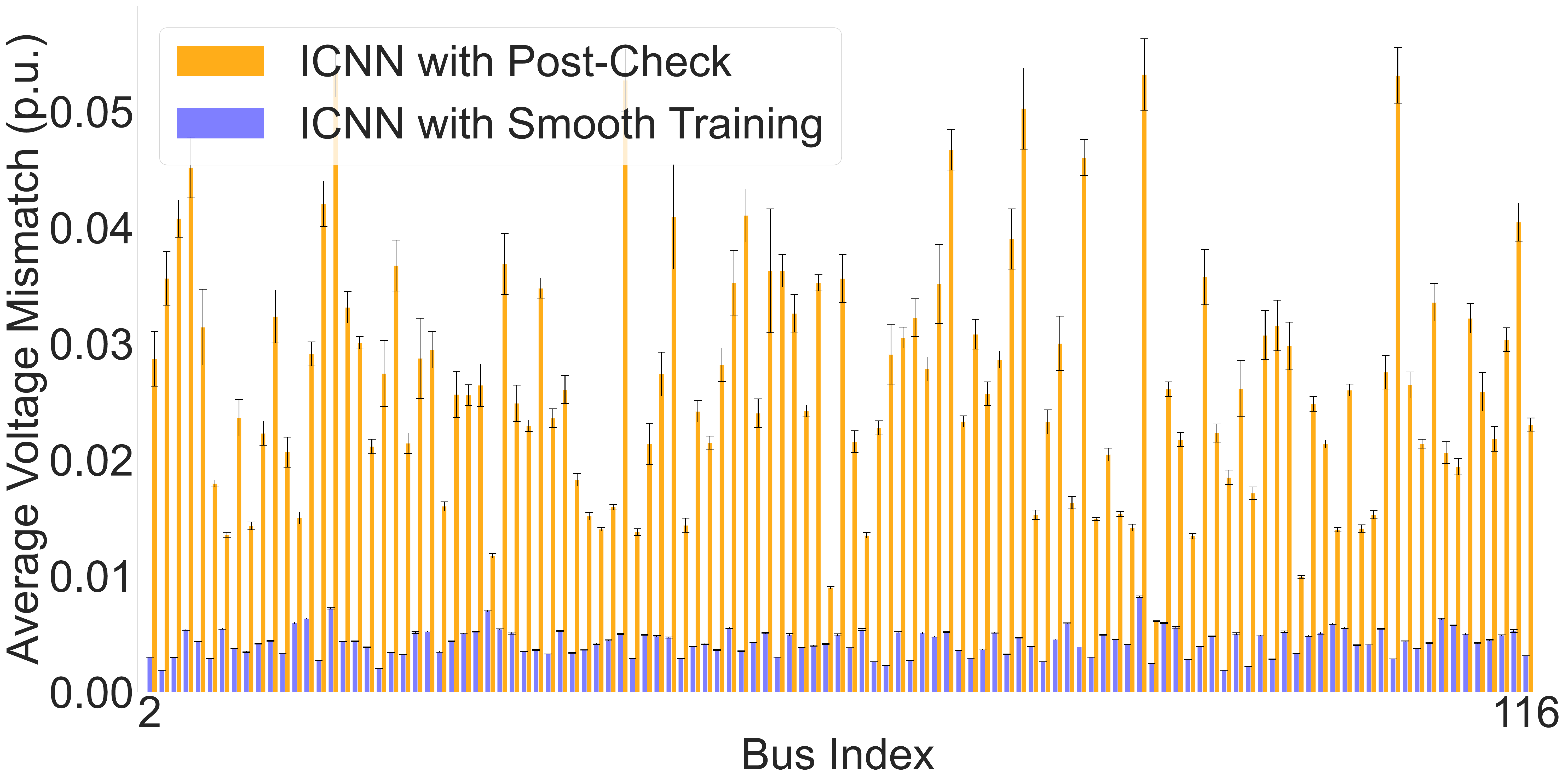}
    \caption{The average voltage mismatches of all experiments on each bus except for the slack bus in the $116$-bus feeder using different training algorithms.}
    \label{fig:average_voltage_mismatch}
\end{figure}

\begin{figure}[h!]
    \centering
    \includegraphics[width=\columnwidth]{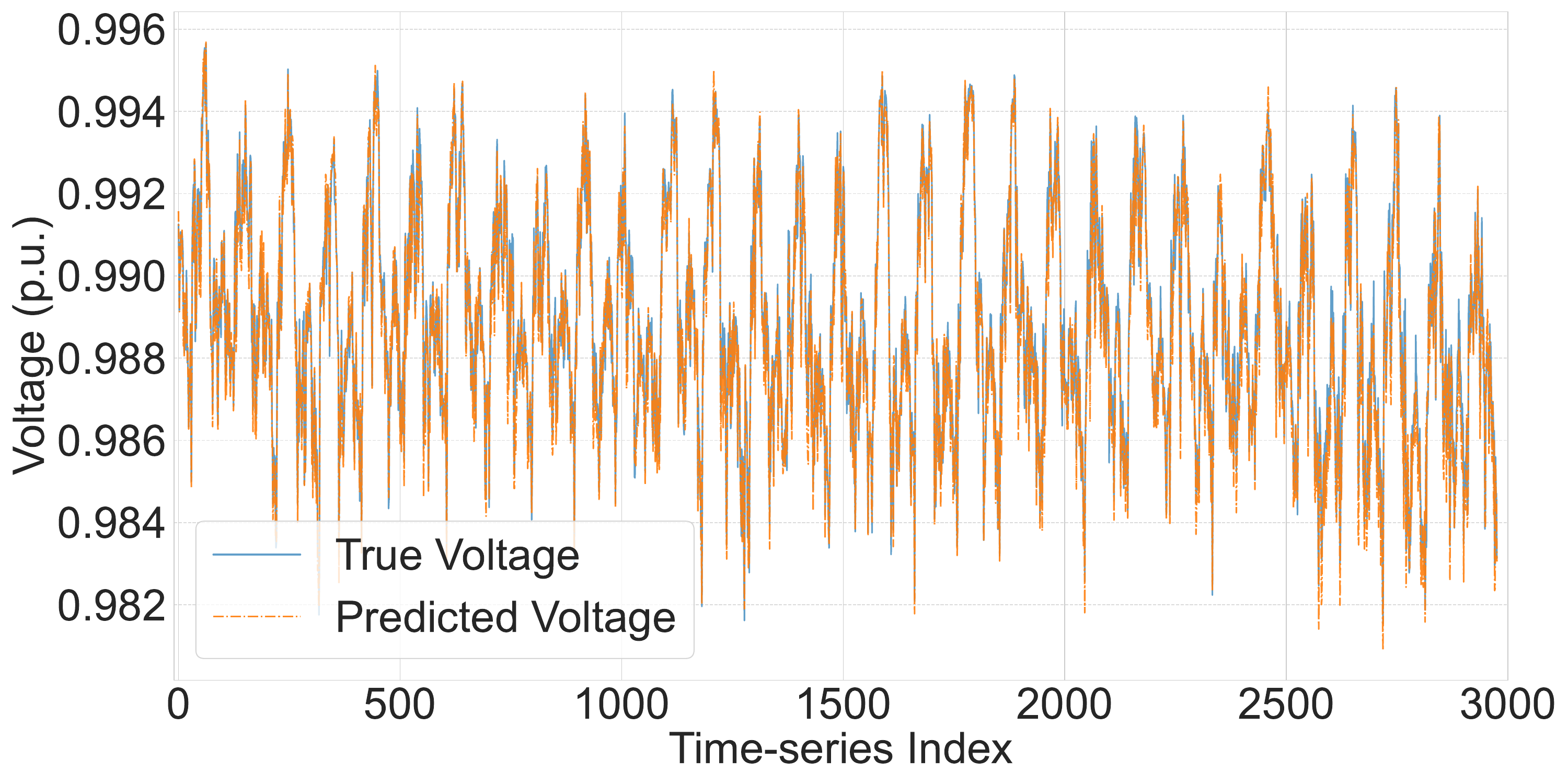}
    \caption{The time-series voltage prediction on one bus in the $116$-bus feeder using the proposed ICNN with smooth training.}
    \label{fig:timeseries}
\end{figure}

For the $116$-bus test case, Fig. \ref{fig:compare_smooth} presents that the ICNN with smooth training exhibits lower training loss and a more robust loss reduction, which is shown as a smaller standard deviation. 
Additionally, the basic ICNN struggles to converge, while the ICNN with the weight gating design achieves a lower loss error. Such observations result from the fact that the parameter initialization or the last update poses a negative value(s) in weights. Using post-processing forces the output weights to be zero when there is a negative weight input.
Specifically, we consider extreme cases for demonstration: initializing with all positive weights for both the basic ICNN and the smooth-trained ICNN yields the same training outcome, which is also the best training performance. The positive weights do not drop significantly during NN updating, so the non-negativity constraint hardly takes effect, and thus has minimal impact on the training. Conversely, the extreme case of initializing with all negative weights leads to a gradient vanishing problem. In this scenario, the weights in $\mW_{1:k-1}$ stop updating due to the gradient vanishing, causing the representation capability to rely heavily on the mapping corresponding with $\mU_{0:k-1}$. 
Despite the extreme cases, we initialized ICNNs with random negative and positive weights to illustrate the general performance of the proposed structure. Moreover, Fig. \ref{fig:average_voltage_mismatch} shows the average predicted voltage mismatch across all experiments for each bus, indicating that ICNN with smooth training has better prediction results. Figure \ref{fig:timeseries} focuses on one bus and explicitly shows the prediction for that bus.


\begin{figure}[h!]
    \centering
    \includegraphics[width=\columnwidth]{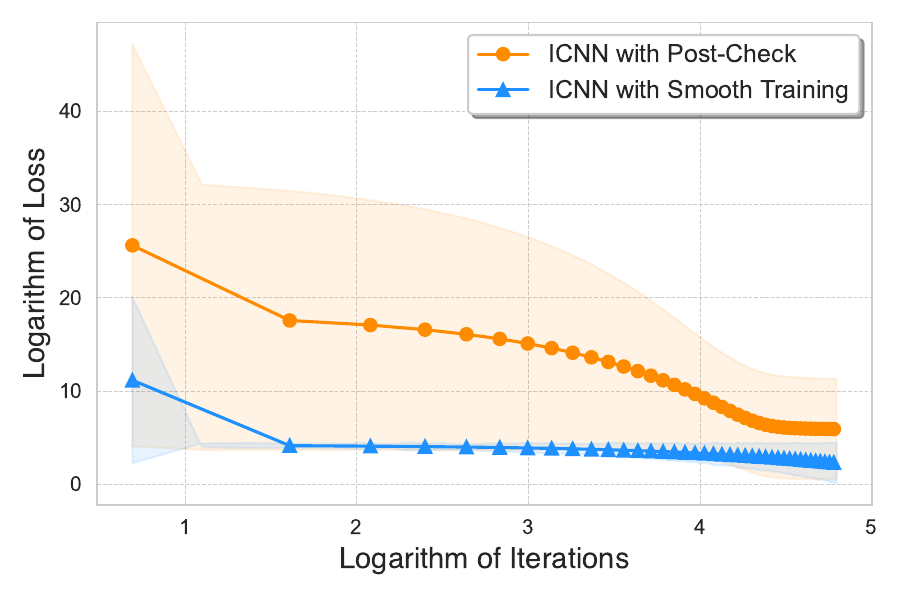}
    \caption{The training error comparison between the basic ICNN and the ICNN with smooth training in the utility feeder test case.}
    \label{fig:power_system}
\end{figure}

Figure \ref{fig:power_system} shows another power system setup in an Arizona generic utility distribution feeder with $371$ buses. But, we notice the same results. The ICNN with integration method for smoothing has a fast convergence speed and better performance. The results indicate that the new design has a robust performance, helping the ICNN better approximate the function $|\mV - \mV^o| = f(\vp, \vq)$, thereby benefiting the finding of the optimal control signal. We also present the bus-level average and maximum MAPE values for all test cases in Table \ref{tab:mape} to provide more detailed training results for all test cases.

\begin{table}[h!]
    \centering
    \begin{tabular}{c|c|c|c|c}
    \toprule
      & \multicolumn{2}{|c|}{ICNN} & \multicolumn{2}{|c}{ICNN}  \\
      & \multicolumn{2}{|c|}{Post-Check} & \multicolumn{2}{|c}{Smooth Training}  \\
     \midrule
     MAPE & Mean & Max & Mean & Max\\
     \midrule
     12-bus & $1.9\%$ & $2.3\%$ & $0.3\%$& $0.4\%$\\
     \hline
     116-bus  & $2.7\%$ & $5.5\%$ & $0.4\%$& $0.8\%$\\
     \hline
     Utility  & $3.3\%$ & $3.7\%$ & $0.9\%$& $1.0\%$ \\
     \bottomrule
    \end{tabular}
    \caption{Comparison of the bus-level mean and maximum MAPE between the basic ICNN and the ICNN with smooth training across various test cases.}
    \label{tab:mape}
\end{table}



\section{Conclusion and Discussion} \label{sec:conclusion}

Our research evaluates existing optimal control methods through data-driven strategies and highlights the limitations of input variable duplication in enhancing the representation power of ICNNs. To address this, we propose an innovative approach that integrates a gate function to impose non-negativity constraints on weights during gradient updates, thereby preserving the model's convexity without disrupting the training dynamics. This approach ensures both minimized loss and the negative coefficient requirement intrinsic to ICNNs. We demonstrate the robustness and applicability of our enhanced ICNN framework through numerical experiments and its successful application to real-world voltage control scenarios. This work not only addresses key challenges in data-driven voltage regulation but also paves the way for future advancements in power system optimization and control. It supports the development of more stable and efficient energy distribution networks, especially in the context of increasing renewable energy integration.

In summary, our work highlights the challenges and opportunities in applying data-driven voltage regulation methods in power grids, especially in scenarios with incomplete or unreliable line connections and parameters. Since the availability of $\vp$ and $\vq$ is also another challenge in the power system, future research will focus on addressing these unobservability issues, ensuring that data-driven approaches can be applied more broadly across different grid conditions.




\bibliographystyle{IEEEtran}
\bibliography{hicss.bbl}

\end{document}